\documentclass[aps,prl,twocolumn,floats,floatfix,superscriptaddress,showpacs]{revtex4-1}
\usepackage{graphicx,amsmath,amssymb,bm,multirow}

\newcommand{\be}[1]{\begin{equation}\label{#1}}
\newcommand{\ee}{\end{equation}}

\newcommand{\bra}[1]{\langle#1|}
\newcommand{\ket}[1]{|#1\rangle}

\begin{document}

\title{Operator evolution for \textit{ab initio} nuclear theory} 
\author{Micah D. Schuster}
\affiliation{San Diego State University, 5500 Campanile Drive, San Diego, California 92182, USA}
\author{Sofia Quaglioni}
\affiliation{Lawrence Livermore National Laboratory, P.O Box 808, L-414, Livermore, 
California 94551, USA}
\author{Calvin W. Johnson}
\affiliation{San Diego State University, 5500 Campanile Drive, San Diego, California 92182, USA}
\author{Eric D. Jurgenson}
\affiliation{Lawrence Livermore National Laboratory, P.O Box 808, L-414, Livermore, 
California 94551, USA}
\author{Petr Navr\'{a}til}
\affiliation{TRIUMF, 4004 Wesbrook Mall, Vancouver, British Columbia, V6T 2A3 Canada}
\pacs{21.60.De, 05.10.Cc, 23.20.Js}

\begin{abstract}
The past two decades have seen a revolution in \textit{ab initio} calculations of nuclear properties.  One key element has been the development of a rigorous effective interaction theory, applying unitary transformations to soften the  nuclear Hamiltonian and hence accelerate the convergence as a function of the model space size. For consistency, however, one ought to apply the same transformation to other operators when calculating transitions and mean values from the eigenstates of the renormalized Hamiltonian.  Working in a translationally-invariant harmonic oscillator basis for the two- and three-nucleon systems, we evolve the Hamiltonian, square-radius and total dipole strength operators by the similarity renormalization group (SRG). The inclusion of up to three-body matrix elements in the $^4$He nucleus all but completely restores the invariance of the expectation values under the transformation. We also consider  a Gaussian operator with adjustable range and find at short ranges an increased contribution from such induced three-body terms.
\end{abstract}

\maketitle

\textit{Ab initio} calculations of atomic nuclei have become increasingly successful in recent years, including first principles calculation of astrophysically relevant fusion reactions \cite{navratil11a, navratil12a}, of the anomalously long lifetime of $^{14}$C \cite{maris11a}, and of the crucial Hoyle state in $^{12}$C \cite{epelbaum11a}.  Of particular importance is the need to include \textit{ab initio} three-body forces, for example, in correctly describing nuclear binding energies and spectra, especially the ground state spin of $p$-shell nuclei \cite{navratil03a}, the lifetime of $^{14}$C \cite{maris11a}, and the location of the neutron drip line for oxygen isotopes \cite{otsuka10a}.

These breakthrough discoveries have been driven by advances in computing, in effective field theory \cite{weinberg90a,weinberg91a,ordonez92a,bedaque02a}, in techniques for the solution of the nuclear many-body problem, such as the no-core shell model (NCSM) \cite{navratil00a,navratil00b, navratil01a}, and in modern effective interaction theory \cite{stetcu12,furnstahl13a}.  The latter takes the form of unitary transformations chosen to reduce the coupling between low- and high-momentum states, which arises from the bare nuclear interaction's ``hard core'' and leads to slow convergence in the size of the model space.

Here we focus on the similarity renormalization group (SRG) \cite{bogner07,bogner10}, which has been successful when computing nuclear properties for a variety of nuclei \cite{stetcu12,furnstahl13a,jurgenson09a,jurgenson11,roth11a,tsukiyamat11a,roth12a,hergert13a}. Independently developed by Glazek and Wilson \cite{glazek93} and Wegner \cite{wegner94}, the SRG is a series of unitary transformations on the Hamiltonian,
\begin{equation}
\hat{H}_s=\hat{U}_s\hat{H}_{s=0}\hat{U}_s^\dagger,
\label{eqn:uhu}
\end{equation}
where $\hat{U}_s$ labels the sequence of transformations starting with the initial Hamiltonian at $s=0$. This can be rewritten as a flow equation in $s$ and an antihermitian generator, $\hat{\eta}_s=d\hat{U}_s/ds ~\hat{U}_s^\dagger$,
\begin{equation}
\frac{d\hat{H}_s}{ds}=[\hat{\eta}_s,\hat{H}_s].
\label{eqn:hflow}
\end{equation}
The generator is commonly chosen to be $[\hat{T}, \hat{H}_s]$, where $\hat{T}$ is the kinetic energy operator. This drives the Hamiltonian towards diagonal form in momentum space, thus weakening the coupling between low- and high-momentum states, though other generators have also been successful \cite{li11}. Rather than use the flow parameter, $s$, it is common to use $\lambda = s^{-1/4}$, to keep track of the sequence of Hamiltonians \cite{jurgenson11}; note that as $\lambda$ decreases, the Hamiltonian will undergo more evolution.

While formally the transformed Hamiltonian should be independent of the unitary transformation and specifically of the SRG flow parameter, the evolution induces higher-order terms, up to A-body, into the Hamiltonian. Previous work has suggested that stopping at three-body terms leads to energies mostly independent of $\lambda$ \cite{jurgenson11}. There is more to physics than energy spectra however.  For instance, we want to accurately quantify electric dipole transitions which lead to important observables that are difficult to measure: e.g., the polarization of a nucleus \cite{stetcu09a}; or the radiative capture $^7$Be$(p, \gamma)^8$B, crucial to understanding the neutrino signature of our sun \cite{adelberger11a,navratil11a}.  When using SRG-evolved Hamiltonians, for consistency one should also evolve any other operator,
\begin{equation}
\hat{O}_s=\hat{U}_s\hat{O}_{s=0}\hat{U}_s^\dagger,
\label{eqn:otrans}
\end{equation}
using the same sequence of unitary transformations that were applied to the Hamiltonian. While this can be rewritten into a similar form as Eq. (\ref{eqn:hflow}), it is more computationally efficient to directly compute the unitary matrix, $\hat{U}_s$ \cite{bogner10},
\begin{equation}
\hat{U}_s=\sum\limits_{\alpha}\ket{\psi_\alpha(s)}\bra{\psi_\alpha(0)},
\end{equation}
where $\ket{\psi_\alpha(0)}$ and $\ket{\psi_\alpha(s)}$ are the eigenvectors of the Hamiltonian before and after SRG evolution, respectively. The transformation of Eq. (\ref{eqn:otrans}) is then given by a simple matrix multiplication.

Evolution of operators is one of the frontiers of renormalization group methods \cite{furnstahl13a} and extending these techniques to any operator is an important goal. A study of M1 and E2 transitions using the Okubo-Lee-Suzuki unitary transformation \cite{stetcu05} shows significant renormalization at the two-body cluster level, especially for short-ranged operators. Previous work using the SRG has focused on how the operators change in momentum space \cite{anderson10}. The work presented in this Letter aims to evolve operators via the SRG, including up to induced three-body terms, and, for the first time, test the consistency of the expectation values as a function of evolution.

We are interested in two areas: (1) dependence of operator expectation values on the SRG parameter, $\lambda$, when applied to SRG evolved wavefunctions, and (2) the effect of range on the amount of renormalization that an operator undergoes. We perform these studies in the three- and four-nucleon systems, where we can obtain accurately converged results for a variety of observables.

We first investigate two observables, the root mean square (RMS) radius of the chosen nucleus, and the total strength of the dipole transition, given by $\bra{\Psi_0}\hat{D}^2\ket{\Psi_0}$, where $\hat{D}$ is the dipole operator,
\begin{equation}
\hat{D}=\sum_{i}^A\left(\frac{1}{2}-\tau_i^z\right)\vec{r}_i,
\end{equation}
and $\ket{\Psi_0}$ is the ground state wavefunction of the nucleus, $\tau^z_i$ is the third component of isospin and $\vec{r}_i$ is the position vector of the $i$th particle.  We choose the total dipole strength because it is used to compute important quantities such as photo-absorption cross sections \cite{quaglioni07} and electric polarizabilities \cite{stetcu09a}. 
Our second investigation focuses on operator renormalization as a function of range, following a prescription similar to that of Ref. \cite{stetcu05}.

Our calculations adopt nucleon-nucleon ($NN$) and three-nucleon ($3N$) forces derived from chiral effective field theory ($\chi$EFT) \cite{epelbaum02a,epelbaum09a} and are performed with the NCSM in a Jacobi harmonic oscillator basis \cite{navratil00b}. This is a translationally-invariant, antisymmetric basis truncated at $N_{\text{max}}\hbar\Omega$ above the lowest many-body configuration, where $\Omega$ is the harmonic oscillator parameter and $N_{\text{max}}$ is the maximum number of excitations. We use ground state wavefunctions calculated from three different Hamiltonians: (1) $NN$-only, two-body Hamiltonian from the SRG evolution of the $NN$ force in the two-nucleon space; (2) $NN$+$3N$-induced, three-body Hamiltonian from the SRG evolution of the $NN$-force in the three-nucleon space; and (3) $NN$+$3N$, SRG Hamiltonian obtained from evolving the $NN$ plus initial $3N$ forces in the three-nucleon system. We construct these Hamiltonians in the same manner as Ref. \cite{jurgenson09a}. The only difference between $NN$+$3N$-induced and $NN$+$3N$ is the inclusion of the initial three-body interaction in the latter, which simply causes an overall shift in the mean values calculated over their eigenstates for $^3$H and $^4$He, similar to that found in energies \cite{jurgenson09a}.

Because we work in relative coordinates, all operators considered here are written as two-body operators. Similar (but not quite parallel) to our three classes of Hamiltonian, we consider three stages of operator evolution: (1) Bare or unevolved operator; (2) 2B evolved, SRG-evolution of the operator in the two-body space; and (3) 3B evolved, SRG-evolution of the operator in the three-body space, allowing the induction of three-body terms.

\begin{figure}[h]
\begin{center}
\includegraphics[trim=0cm 0.0cm 0.0cm 0.05cm,width=0.45\textwidth]{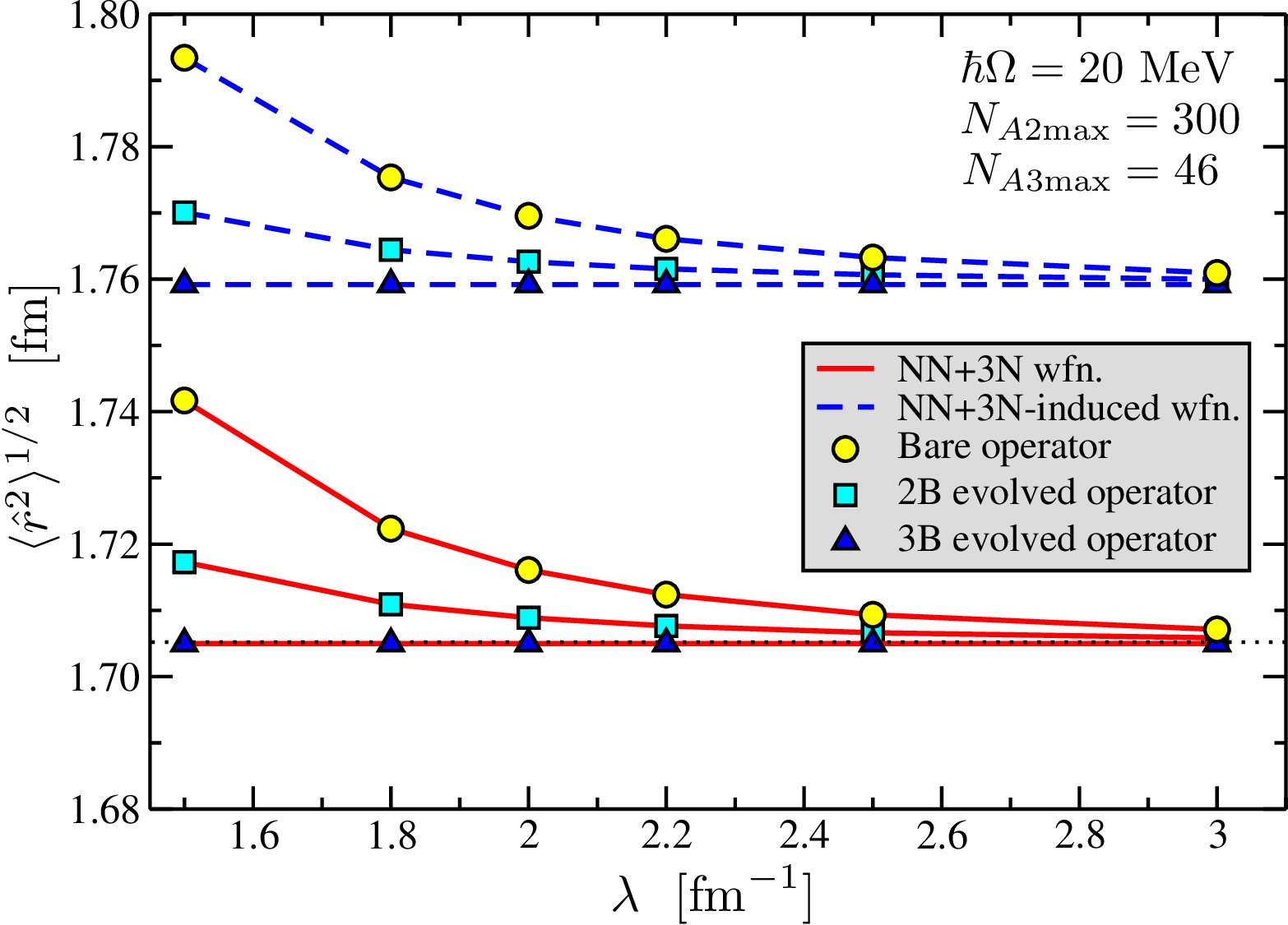}
\end{center}
\caption{(color online). $^3$H RMS radius as a function of SRG evolution parameter, $\lambda$. Shown are results obtained with wavefunctions from two Hamiltonians: $NN$+$3N$-induced (blue dashed line) and $NN$+$3N$ (red solid line), and three levels of operator evolution: bare operator (circles), operator evolved in the two-body space (squares), and operator evolved in the three-body space (triangles). The dotted line is the RMS radius calculated using the bare Hamiltonian and bare operator.}
\label{fig:3H}
\end{figure}

We first verified that the two- and three-body SRG transformations of external operators are unitary in the two- and three-nucleon systems, respectively. To this end we calculated the expectation value of the renormalized $\hat{r}^2$ operator  on the ground state wavefunctions of $^2$H and $^3$H nuclei. Fig. \ref{fig:3H} shows the RMS radii of $^3$H for the three levels of operator evolution described previously, with a range of $\lambda$ from $1.5$ fm$^{-1}$ to $3.0$ fm$^{-1}$ and $\hbar\Omega=20$ MeV. This range of $\lambda$ has shown to improve convergence for energy calculations \cite{jurgenson09a}. To obtain converged expectation values (to within less than 0.1\%) we truncate the $A=2$ model space at $N_{\text{max}}=300$ and the $A=3$ model space at $N_{\text{max}}=46$, denoted as $N_{\text{A2max}}$ and $N_{\text{A3max}}$, respectively. 

As expected, when using the bare operator the RMS radius has a clear dependence on $\lambda$ even when the Hamiltonian includes three-body SRG induced terms. When the operator is evolved in the two-body space, the dependence is reduced but still significant. However, when evolved in the three-body space there is no dependence on $\lambda$ because both the Hamiltonian and the $\hat{r}^2$ operator include all SRG induced terms, thus the transformation is exactly unitary. 
This is confirmed by the agreement with 
the expectation value calculated using the bare Hamiltonian and bare operator, 
shown in Fig. \ref{fig:3H} as a dotted line.

\begin{figure}[h]
\begin{center}
\includegraphics[trim=0cm 0.0cm 0.0cm 0cm,clip,width=0.4\textwidth]{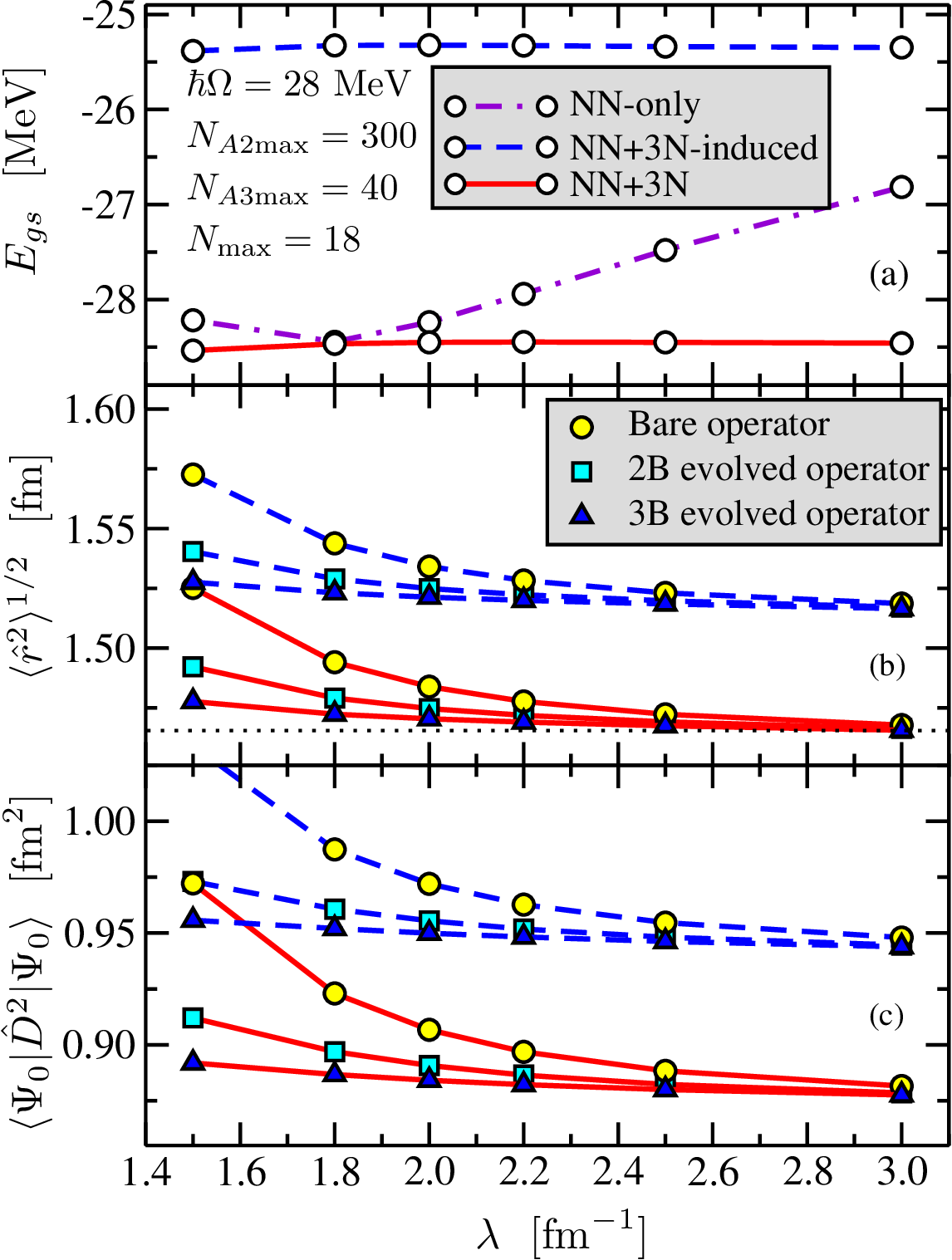}
\end{center}
\caption{(color online). Calculations of $^4$He ground state energy (a), RMS radius (b), and total strength of the dipole transition (c) for $\text{N}_\text{max}=18$, with a range of $\lambda$ from 1.5 fm$^{-1}$ to 3.0 fm$^{-1}$. The purple dot-dashed line indicates results obtained with the $NN$-only Hamiltonian. The dotted line indicates the expectation value computed using the bare Hamiltonian and bare operator. See also caption of Fig. 1}
\label{fig:4He}
\end{figure}

We next extend these calculations to $^4$He, and compute the RMS radius and total strength of the dipole transition. Fig. \ref{fig:4He} shows these three calculations in a range of $\lambda$ from $1.5$ to $3.0$ fm$^{-1}$ with $\hbar\Omega = 28$ MeV. We truncate the $A=2$ model space at $N_{\text{A2max}}=300$, the $A=3$ model space at $N_{\text{A3max}}=40$, and the the $A=4$ space at $N_{\text{max}}=18$ which leads to converged results within less than $0.1\%$ for both observables. Results for the ground state energy, in this range of $\lambda$, have been studied in detail previously \cite{jurgenson11}. We show them here, panel (a), to emphasize that when one does not include SRG induced three-body terms into the Hamiltonian, ($NN$-only curve), the ground state energy is dependent on $\lambda$ over the entire range we investigate. However, when the three-body terms are included, the ground state energy is independent of $\lambda$ above $1.8$ fm$^{-1}$. Below $\lambda=1.8$ fm$^{-1}$ the binding energy drops due to the missing four-body SRG induced terms. 

Panels (b) and (c) show the results for the RMS radius and total strength of the dipole transition, respectively. The trends of these results are similar because the operators are closely related \cite{quaglioni07}. When using the bare operator, the observable has a significant $\lambda$ dependence at small values. However, when evolved in the two- and then in the three-body space, independence is all but restored. The transformation is not completely unitary due to the SRG induced four-body terms that we do not include. This causes a slight increase in the 
calculated observables at smaller $\lambda$ values, as emphasized for the RMS radius, 
for which we show also the expectation value obtained with the bare $NN$+$3N$ Hamiltonian and bare operator (dotted line).  
This bare result can be also recovered at large lambda values, where the induced terms affecting the operator become increasingly smaller. The trade off, however, is a much slower convergence rate, which would require prohibitively large model space sizes for heavier-mass systems. There, $\lambda$ is typically chosen between $1.8$ and $2.0$ fm$^{-1}$, where one can speed up the convergence while keeping to a minimum the effect of beyond-three-body induced forces.

\begin{figure}[ht]
\begin{center}
\includegraphics[trim=0cm 0.0cm 0.0cm -0.1cm,width=0.45\textwidth]{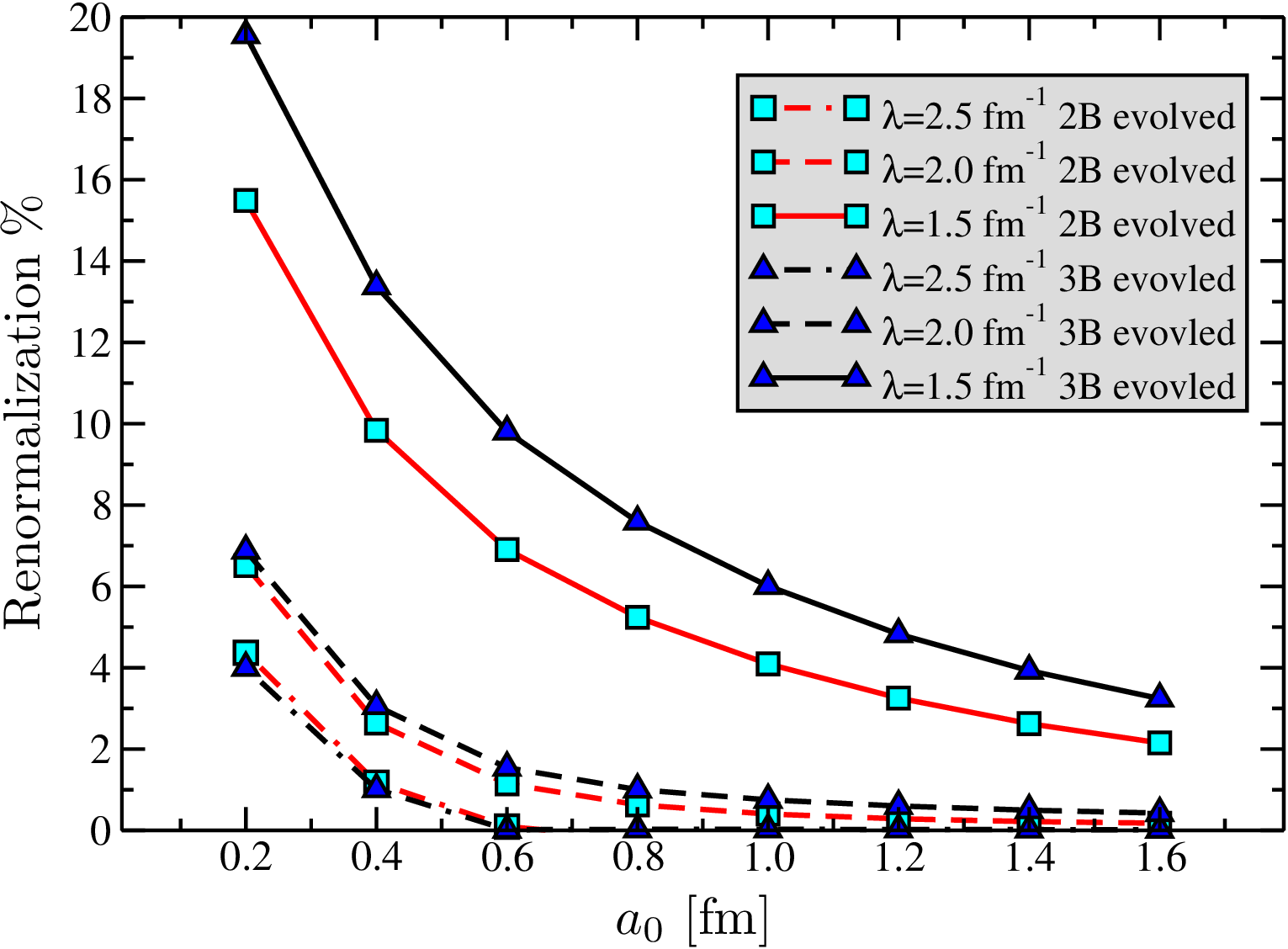}
\end{center}
\caption{(color online). Renormalization percent as a function of range of a Gaussian operator, see Eq. (\ref{eq:gau}), for three values of the SRG parameter, $\lambda$: $1.5$ fm$^{-1}$(solid line), $2.0$ fm$^{-1}$(dashed line), and $2.5$ fm$^{-1}$(dot-dashed line). Symbols as in the caption of Fig. \ref{fig:3H}. The wavefunction is from $NN$+$3N$ SRG evolved Hamiltonian.}
\label{fig:gau}
\end{figure}

Our investigation so far has considered two long range operators, $\hat{r}^2$ and $\hat{D}^2$, and has shown a relatively small, but non-trivial, renormalization. To highlight the importance of operator range when using the SRG method, in combination with operators evolved in the three-body space, we use a Gaussian two-body operator of range $a_0$, 
\begin{equation}
\hat{\text{O}}(\vec{r}_1,\vec{r}_2)=A~\text{exp}\left(-\frac{(\vec{r}_1-\vec{r}_2)^{~2}}{a_0^2} \right),
\label{eq:gau}
\end{equation}
where $A$ is the normalization chosen to be
\begin{equation}
A\int \text{exp}\left(-\frac{r^2}{a_0^2} \right)d\vec{r}=1.
\end{equation}
This follows a similar prescription to that of Ref. \cite{stetcu05}, where the authors focus on operator range and Okubo-Lee-Suzuki renormalization. Although this operator does not represent any physical phenomena, one can easily adjust its range, giving us a systematic way to explore the amount of renormalization for operators evolved via SRG. We define the amount of renormalization as $(\langle\hat{\text{O}}_{\text{eff}}\rangle-\langle\hat{\text{O}}_{\text{bare}}\rangle)/ \langle\hat{\text{O}}_{\text{bare}}\rangle \times 100$. Fig. \ref{fig:gau} shows the results using the same $^4$He ground state wavefunction as above.

At short ranges the expectation values computed with the SRG evolved operator, whether evolved in the two- or three-body space, are significantly renormalized from the bare operator, while as the range increases, the renormalization tends towards zero. More interesting is the three-body contribution to the overall renormalization, i.e. the difference between the expectation values of the operator evolved in the two-body space and that of the operator evolved in the three-body space for the same value of $\lambda$. The relative three-body contribution tends to increase as the range increases, approximately 25\% at $a_0=0.2$ fm to 50\% at $a_0=1.6$ fm, even though the absolute magnitude of the three-body contribution decreases.  Beyond $\lambda=2.5 \text{ fm}^{-1}$, the renormalization percent is close to zero for all but the shortest ranges, so we do not show larger values of $\lambda$ here. This shows that the amount of renormalization that occurs to an operator is highly dependent on that operator's range, confirming, and extending on, previous work done in the two-body space \cite{stetcu05,anderson10}.

In summary, we have, for the first time, SRG evolved several operators in the two- and three-body spaces and computed expectation values using ground state wavefunctions of $^3$H and $^4$He.  For $A=3$ this completely restored unitarity, that is, independence of the SRG evolution parameter $\lambda$. Including up to three-body induced terms in the $A=4$ system, the dependence on $\lambda$ was dramatically reduced, but not eliminated, due to the induced four-body terms. By using a Gaussian operator with adjustable range, we demonstrated the relative size of the induced three-body terms were larger for shorter ranges. Future work will include adding the ability to evolve non-scalar operators, which will allow us to investigate other quantities such as transition strengths and cross-sections. We will also extend these calculations to heavier systems (e.g. $A=5$ - $12$), where it is computationally more advantageous to work with single-particle Slater determinant basis states. This can be accomplished by transforming the present translationally invariant three-body operators into matrix elements over Slater determinant three-nucleon basis states, similarly to what has been done for the $3N$-force. 

\acknowledgments

This work was performed under the auspices of the U.S. Department of Energy by Lawrence Livermore National Laboratory under Contract DE-AC52-07NA27344. 
Support came from U.S. DOE/SC/NP (work proposal SCW1158), U.S. Department of Energy grants DE-FG02-96ER40985 and DE-FC02-07ER41457 and the Natural Sciences and Engineering Research Council of Canada (NSERC) Grant No. 401945-2011. TRIUMF receives funding via a contribution through the National Research Council Canada. Computing support came from the LLNL institutional Computing Grand Challenge program.

\end{document}